\documentclass{elsart3p}

 \usepackage{epsfig}

\usepackage{amssymb}

\begin{document}

\begin{frontmatter}


 \title{Systematic uncertainties in MonteCarlo simulations of the atmospheric muon flux in the 5-line ANTARES detector}
\author[a]{Annarita Margiotta\corauthref{cor1}}
 \ead{margiotta@bo.infn.it}
 \corauth[cor1]{on behalf of the ANTARES collaboration.\\ Prepared for the International Workshop on a Very Large Volume $\nu$ Telescope for the Mediterranean Sea, 22-24 Oct. 2007, Toulon, France.}


 \address[a]{Dipartimento di Fisica dell'Universit\`a and Sezione INFN di Bologna, \\
 viale C. Berti-Pichat,6/2, 40127 - Bologna, Italy}



\begin{abstract}
The ANTARES detector was operated in a configuration with 5 lines for a period of 10 months from February until November 2007. The duty cycle was better than 80\% during this period and almost $2\cdot10^7$ atmospheric muon triggers were collected. This large sample was used to test Monte Carlo simulation programs and to evaluate possible systematic effects due to uncertainties on environmental parameters and detector description.  First results are presented and discussed.
\end{abstract}

\end{frontmatter}

\section{Introduction}
\label{sec:Introduction}
The ANTARES Collaboration has just completed the construction of a neutrino telescope off the Southern French coast, 40 km from Toulon, at a depth of about 2500 m under the sea level. Technical details on the ANTARES detector and first results are presented and discussed at this workshop \cite{circella}. 
The main goal of ANTARES is the detection of high energy neutrinos of astrophysical origin. Nevertheless, the most abundant signal is due to high energy muons remaining from the extensive air showers produced in interactions between primary cosmic radiation and atmospheric nuclei. The shielding effect of the sea reduces their flux, which, however, is several orders of magnitude larger than the atmospheric neutrino flux. They represent a dangerous background for track reconstruction as their Cherenkov light can mimic fake upward going tracks. On the other hand they are a useful tool to test offline analysis software, to check our understanding of the detector and to estimate systematic uncertainties.\\
From February to November 2007, the detector ran in a 5-line configuration and collected almost $2\cdot 10^7$ events mainly due to atmospheric muons. This large amount of data was  used for comparisons with the results from a Monte Carlo (MC) simulation. Moreover, several MC samples were produced, using different sets of input parameters, to give an estimate of the systematic effects due to environmental and geometrical parameter uncertainties. 
\section{Monte Carlo simulation}
\label{sec:MonteCarlo}
Two different approaches can be adopted to simulate the atmospheric muon flux: 
\begin{itemize}
	\item a parameterized description of underwater muon flux,
	\item a full Monte Carlo simulation.
\end{itemize}
In ANTARES software both methods are used.
The former is based on the MUPAGE program, described at this conference in \cite{spurio}. \\
In the present analysis the second approach is used for comparison to data and to evaluate systematic errors. In the following its general scheme is presented.\\
The first step of the full MC simulation consists of the generation of a large number of air showers induced by primary nuclei with energy ranging from 1 to $10^5$ TeV/nucleon and zenith angles between $0^o$ and $85^o$
using the CORSIKA software (vrs.6.2) \cite{corsika} 
and the hadronic interaction model QGSJET.01c \cite{qgsjet}. The energy spectrum used for generation is $E^{-2}$. 
Then, all muons reaching the sea level, with energy larger than a threshold energy, are propagated through sea water  to the detector, using MUSIC (MUon SImulation Code) \cite{music}, a 3D muon propagation program that takes into account the main energy loss mechanisms for muons.\\
Finally, muons are transported through the ANTARES sensitive volume, Cherenkov light is produced and the detector response is simulated. At this point, files containing information on hit photomultipliers (PMTs) are translated into the same format used for data, after the addition of a background chosen by the user and a trigger selection, and are ready to be treated with the same reconstruction and analysis program chain used for data. 
The composition of the primary cosmic ray flux is defined by the user. In this study,  a simplified version (only 5 mass groups are considered) of the model presented in \cite{horandel} (Horandel model) is used.
\section{Systematic effects}
\label{sec:SystematicEffects}
\begin{figure}
  \begin{center}
  \mbox{\includegraphics[width=8.5cm]{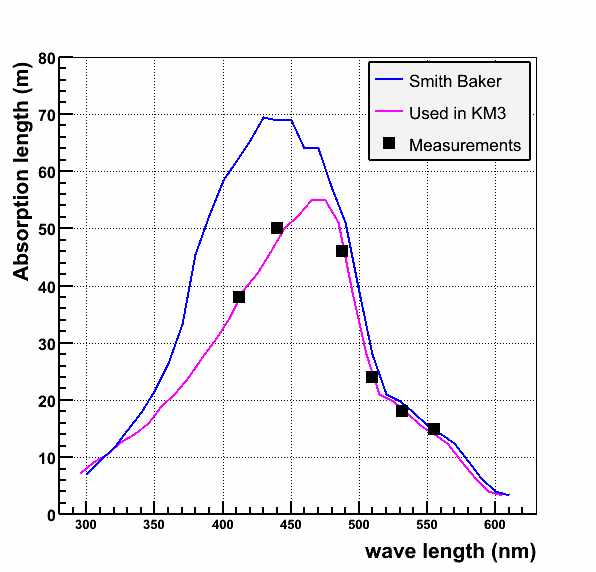}}
\caption{\label{fig1}\footnotesize {Absorption length used as input for ANTARES MC simulation. The lower line is a fit to measurements (black points) taken in the ANTARES site. The upper line is the absorption length for pure water and is shown for comparison.}}
\end{center}
\end{figure}

\begin{figure}
  \begin{center}
  \mbox{\includegraphics[width=8.5cm]{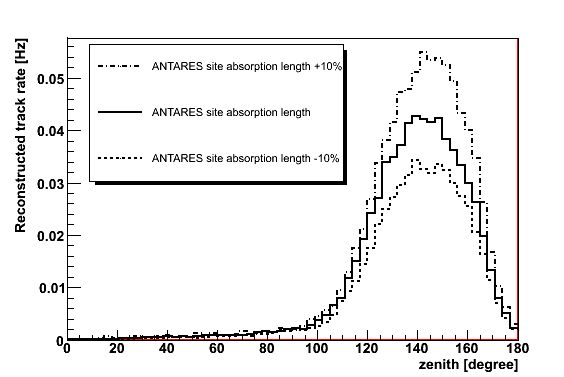}}
\caption{\label{fig2}\footnotesize {Zenith distributions of reconstructed tracks using different absorption lengths. The upper (lower) histogram corresponds to an increase (decrease) of the absorption length by 10 \% wrt the reference values.}}
\end{center}
\end{figure}

\begin{figure}
  \begin{center}
  \mbox{\includegraphics[width=4.5cm]{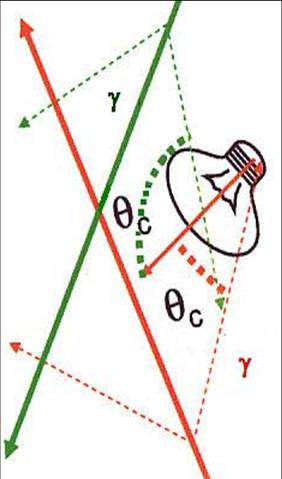}}
\caption{\label{fig3}\footnotesize {Description of the angle $\vartheta_c$ used to define the limit of detection of Cherenkov photons. Two cases corresponding to a downward going and to an upward going  muon are shown. }}
\end{center}
\end{figure}

\begin{figure}
  \begin{center}
  \mbox{\includegraphics[width=8.5cm]{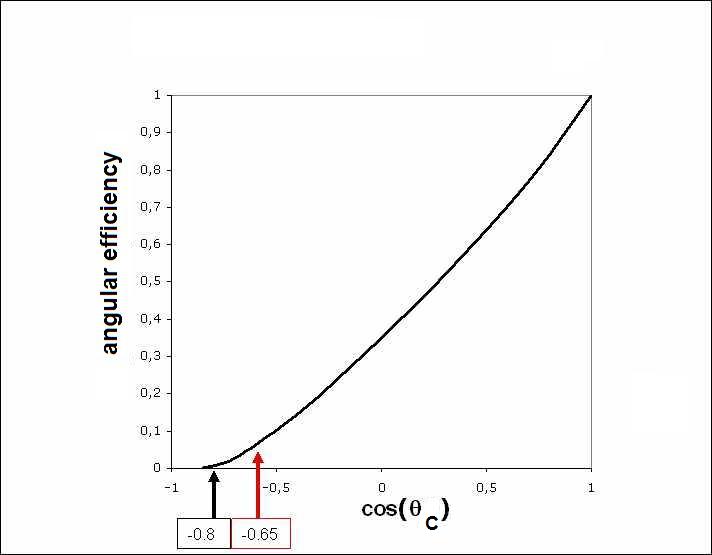}}
\caption{\label{fig4}\footnotesize {Parameterization of OM angular efficiency with the indication of the two different cutoff values used in the systematic effect study.}}
\end{center}
\end{figure}

\begin{figure}
  \begin{center}
  \mbox{\includegraphics[width=8.5cm]{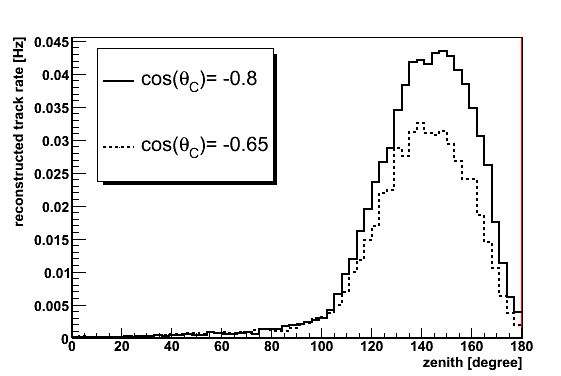}}
\caption{\label{fig5}\footnotesize {Effect of the angle $\vartheta_c$ cut-off on zenith distribution of reconstructed tracks: solid  line refers to cos($\vartheta_c$)=-0.8 (reference value in MC simulation), dotted line represents cos($\vartheta_c$)=-0.65.}}
\end{center} 
\end{figure}

\begin{figure}
  \begin{center}
  \mbox{\includegraphics[width=8.5cm]{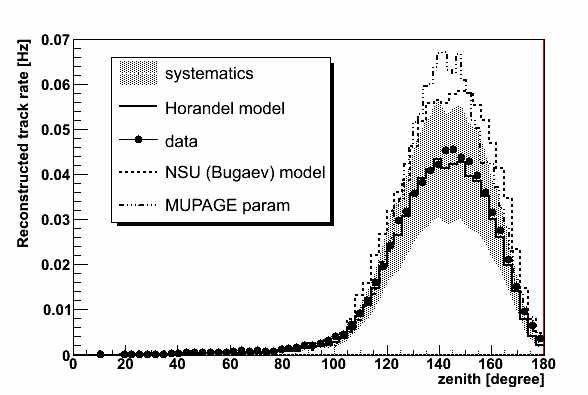}}
\caption{\label{fig6}\footnotesize {Zenith distribution of reconstructed tracks. Black points represent data. Lines refer to MC expectations:  solid line = full simulation with CORSIKA code + QGSJET01 for hadronic interaction description and Horandel model, \cite{horandel};  dotted  line = full simulation with CORSIKA code + QGSJET01 for hadronic interaction description and NSU model, \cite{nsu}; dashed-dotted  line = MUPAGE parameterization, \cite{spurio}. The shadowed  band represents the systematic error, starting from the Horandel model, due to environmental and geometrical parameters. See text for details.}}
\end{center}
\end{figure}

\begin{figure}
  \begin{center}
  \mbox{\includegraphics[width=8.5cm]{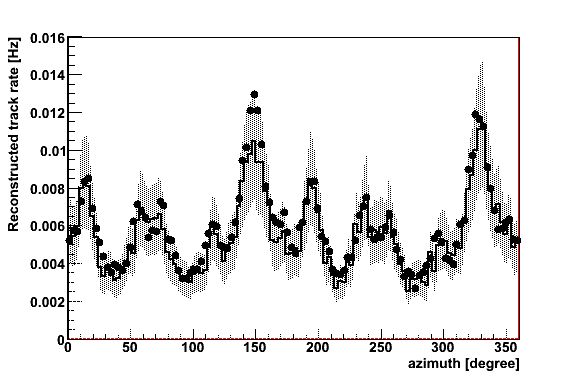}}
\caption{\label{fig7}\footnotesize {Azimuth distribution of reconstructed tracks. Black points represent data. The line refers to MC expectations obtained from the full simulation with CORSIKA code + QGSJET01 for hadronic interaction description and Horandel model. The shadowed  band represents the systematic error due to environmental and geometrical parameters. See text for details.}}
\end{center}
\end{figure}

During MC simulation several input parameters are required to define the environmental and geometrical characteristics of the detector. Some of them play a role as sources of systematic uncertainties.\\
In the following the effect of water absorption length and of PMT description on the reconstructed track rate is considered.\\ 
The absorption length of light in water depends on the light wavelength and has been measured during several sea campaigns. The reference values used in the MC simulation are the result of a fit to a set of measurements collected in the ANTARES site, fig. \ref{fig1}. 
Varying by $\pm 10\%$ the reference values of absorption length, 2 new samples of atmospheric muons were created and compared to the standard production. The 3 zenith distributions are shown in fig. \ref{fig2}. There is an almost negligible effect on the shape of  the histograms, while the absolute flux changes by +25\%/-20\%.\\ 
The geometrical description of the photomultiplier implies the knowledge of its effective area, which is the result of a convolution of 
several factors like quantum efficiency, geometrical surface of the photocathode and collection efficiency. The standard MC simulation
takes as reference values what is reported on Hamamatsu technical sheets.  A sample of simulated muons has been prepared using a PMT efficiency
decreased by 10\%, considering the official Hamamatsu values as the most optimistic set of parameters. A decrease of about 15\% was
observed in the muon flux. Decreasing the effective area by the same amount produces a reduction of 7\% of the atmospheric neutrino flux.\\
Finally, the effect of the maximum angle  between the PMT axis and the Cherenkov photon direction ($\vartheta_c$, see fig. \ref{fig3}) allowing light collection has been studied. 
In fig. \ref{fig4} the OM efficiency is shown as a function of $\vartheta_c$.
Presently, a cut-off at cos($\vartheta_c$)=-0.8 is used. It means that for $\vartheta_c$ larger than $\sim 140^o$ no photon detection is expected. Moving the cut-off to -0.65 (corresponding to $\vartheta_c \sim 130^o$) a decrease of the rate of reconstructed tracks of  about 20\% is observed, see fig. \ref{fig5}. \\ 
Given PMTs' orientation (looking downward at 45$^o$  with respect to the vertical) the effect on neutrino detection is lower.\\
Summing in quadrature the different contributions, a global systematic effect of about $\pm 30\%$ can be considered as an estimate of the errors produced by uncertainties on environmental and geometrical parameters.  New measurements are in progress both of absorption length, using in-situ tools, and of geometrical characteristics of the PMTs and a more precise definition of the actual values is expected.\\
Another source of uncertainty, which is common to all astroparticle physics experiments, is represented by the choice of the primary cosmic ray composition and of hadronic interaction description.
Some experimental measurements suggest that the Horandel model \cite{horandel}, adopted in this MC simulation, underestimates the all-particle spectrum by about 30\% (see, for example \cite{bugaev} and \cite{kascade}). \\
A similar uncertainty can be considered as an estimate of the systematic effect due to the choice of the hadronic interaction model. \\
To give an idea of the possible difference among different parameterizations of primary cosmic ray composition, in fig. \ref{fig6} the zenith distribution of data (black points) is compared to MC expectations obtained using the Horandel model (solid line) \cite{horandel} , the NSU model (dotted line) \cite{nsu}  and the MUPAGE parameterization (dashed-dotted line) \cite{spurio}. The shadowed band represents the systematic error due to environmental and geometrical parameters. In fig. \ref{fig7} the azimuth distribution is shown. Only data and MC with Horandel model are shown, together with the systematic error band. The pictures show a good agreement between data and MC both in shape and in absolute normalization, within the estimated uncertainties. 
\section{Conclusions}
\label{sec:Conclusions}
A comparison between data and results from MC simulation of atmospheric muon flux is presented. The distributions of zenith and azimuth show a good agreement both in shape and in the absolute flux. \\
The effect of environmental and geometrical parameter uncertainties on reconstructed track rate is studied. Presently, a value of about $\pm 30\%$ can be considered as a first evaluation of the systematic errors. New measurements are in progress to reduce uncertainties on the parameters used as input for MC simulations.  The systematic effect on atmospheric neutrino flux predictions is noticeably lower.\\


\end{document}